# Improving Power Grid Resilience Through Predictive Outage Estimation

Rozhin Eskandarpour, Amin Khodaei
Department of Electrical and Computer Engineering
University of Denver
Denver, CO 80210, USA
rozhin.eskandarpour@du.edu, amin.khodaei@du.edu

Ali Arab
Data Management and Advanced Analytics
Protiviti Inc.
New York City, NY 10019, USA
ali.arab@protiviti.com

*Abstract—* In this paper, in an attempt to improve power grid resilience, a machine learning model is proposed to predictively estimate the component states in response to extreme events. The proposed model is based on a multi-dimensional Support Vector Machine (SVM) considering the associated resilience index, i.e., the infrastructure quality level and the time duration that each component can withstand the event, as well as predicted path and intensity of the upcoming extreme event. The outcome of the proposed model is the classified component state data to two categories of outage and operational, which can be further used to schedule system resources in a predictive manner with the objective of maximizing its resilience. The proposed model is validated using *k*-fold cross-validation and model benchmarking techniques. The performance of the model is tested through numerical simulations and based on a well-defined and commonly-used performance measure.

I.  INTRODUCTION

Predictive analytics and emerging applications of machine intelligence tools are shaping every aspects of our daily lives. Data has become the epicenter of the modern decision making by policy makers, corporations, and enterprises. Utilities and local governments are facing increasing expectations from their customers and constituencies to effectively respond to the aftermath of the catastrophic events such as hurricanes that can affect the quality of life of the communities and interrupt the business continuity. In this climate, the concept of resilience enhancement has become an important risk management measure in addressing these challenges.

Resilience denotes the capability of a system to absorb and to adapt to external shocks, which is an important characteristic expected from critical lifeline systems such as electric power grids [1]. There are several types of external shocks to the power grid, most notably extreme events which include adverse weather events and natural disasters that are known to cause considerable negative impacts not only on the system itself but also on the society in general [2]. Among these extreme events, hurricanes are known to be the most frequent extreme event in the United States, mainly occurred along the Atlantic Ocean and Gulf of Mexico [2]. The devastating aftermath of these events calls for disruptive strategies to ensure that the power grid can still supply electricity to customers, or even if considerably impacted, can quickly bounce back from the contingency state to its normal operational condition. In this case, an accurate forecasting of the likely hurricane impacts on the power grid can be of significant value as it can be leveraged in achieving enhanced grid resilience. This paper proposes a machine learning based method for predicting the state of the power grid components in response to upcoming hurricane strikes.

The concept of resilience for complex systems was originally introduced by Holling [3] in the ecology area. Holling defined the resilience of a system as the rate and speed of returning to normal conditions after an extreme event. The intent of resilience study is to anticipate the unexpected change due to failure, considering that systems have limits and gaps, and the atmosphere constantly affects both regarding design and external shocks [4]. In [5], the significance of geographic and cascading interdependencies are highlighted which are associated with urban infrastructure, and a general method to describe infrastructure interdependencies is proposed. In [6] the impact of resilient systems on diminishing the probabilities of failure in urban infrastructure is analyzed. This concept was extended into other systems including the power grids. In [7] an approach for calculating the resilience of a single infrastructure and its components is proposed. In [8] a proactive resource allocation method aiming to repair and recover power grid after extreme events is proposed. In [9] and [10] a proactive recovery framework of power grid components is introduced which develops a stochastic model for operating the components prior to the event, followed by a deterministic recovery model for managing resources after the event. In [11] a restoration model is proposed based on power flow constraints which identifies an optimal schedule using the macroeconomic concept of the value of lost load (VOLL) in order to minimize the economic loss due to load interruptions in the post-disaster phase. A decision-making model, based on unit commitment solution and system configuration, is proposed in [12] to find the optimal repair schedule after a hurricane and in the restoration phase of a damaged power grid.

This work has been supported in part by the U.S. National Science Foundation under Grant CMMI-1434771.

In [13], a power grid resilience index is proposed by analyzing the process of generation, transmission, and consumption of electricity in various countries. The geometric mean of several factors such as the generation efficiency of non-renewable fuel dependence, the distribution efficiency, the carbon intensity, and the diversity are considered to develop the resilience index. However, an index for individual components in the system is not considered in the methodology. In [14], a methodology to calculate resilience index of power delivery systems in post-event infrastructure recovery is proposed. A multi-infrastructure system including electric power delivery, telecommunications, and transportation is considered and the resilience measures of fragility and quality are combined with the input-output model of these infrastructures. The proposed index is evaluated by the data collected from post-landfall of Hurricane Katrina to assess the resilience and interdependence of a multi-system networked infrastructure during natural extreme events. The study in [15] proposes a framework for resilience enhancement of urban infrastructure systems. The time-dependent expected resilience metric is built on performance and response of the power grid following an extreme event. The process is performed in the stages of disaster prevention, damage propagation, and assessment and recovery. The hurricane resilience of electric power grids is quantified through a probabilistic modeling approach in [16], using a Poisson process model for hurricane occurrence, component fragility models, and a grid restoration model with component repair priority. The model is then calibrated using actual customer outage and power grid restoration data in Harris County, Texas in the aftermath of Hurricane Ike in 2008.

In many problems, a closed formulation of the problem and its solution cannot be easily derived. Machine learning investigates the algorithms that are capable of learning from and making forecasts from data. These algorithms can categorize the observed data for classification (supervised learning), combine similar patterns for clustering (unsupervised learning), and predict the output of the system based on its past behavior and historical data (regression modeling) [17]. Machine learning approaches have been utilized in a considerable number of research efforts in the power and energy sector [18]. Security assessment is one of the most versatile machine learning applications in power grids with the applications from pattern recognition [18], decision tree induction, and nearest neighbor classifiers [19], to name a few. Forecasting arises as another popular application of machine learning. A number of Artificial Neural Networks (ANNs) have been proposed for short-term load forecasting [20] and wind power forecasting [21]. Some other examples of machine learning applications in power grids include risk analysis using regression models, ANNs, and Support Vector Machine (SVM) [22], distribution fault detection applying ANNs and SVM [23], and power outage duration prediction using regression models and regression trees/splines [24].

This paper proposes a three-dimensional SVM [25] model to predict the potential power grid component outages in response to hurricanes. The proposed SVM is trained on historical data with three features, namely the resiliency index of the component, the distance of the component from the center of the hurricane, and the category of the hurricane which is determined based on the wind speed. The rest of the paper is organized as follows: Section II presents the model outline and formulation of the proposed machine learning method for outage prediction. Section III presents simulation results on a test system, and Section IV concludes the paper.

## II. PROBLEM STATEMENT

Resilience index is measured by considering the quality of a component and its performance in a time period before and after extreme events. The state of a component during a hurricane can be considered as *damaged* (component is on outage) or *operational* (component is in service). In order to classify the damage/operational state of each component, various features borrowed from historical data can be used. In this paper, a three-dimensional SVM is utilized in order to determine the state of the components and the decision boundary between the damaged and operational data points. In the following, a brief introduction on multi-dimensional SVM is provided and the features that are used to determine the state of the components are discussed.

### A. Support Vector Machines

SVM is a discriminative classifier that defines a separating hyperplane between two classes. The best hyperplane in SVM is considered as the hyperplane with the widest gap between the classes which decreases the risk of miss-classifying and increases the generalization of the classifier. This gap is usually referred to as margin, where SVM intends to maximize this margin between the classes.

The details of the SVMs are fully described in the literature [25], so only a brief introduction to SVM in three-dimensional space is presented in this section. Consider $m$ training samples $x_i \in R^3$, $i=1,...,m$ in a binary classification problem. The linear decision is function $f(x)=\text{sign}(w^T x+b)$, $x_i \in R^3$, where $w$ is the weight vector which defines a direction perpendicular to the hyperplane of the decision function, while $b \in R$ is a bias which moves the hyperplane parallel to itself. The optimal decision function given by support vectors is the solution of the following optimization problem:

$$\min \frac{1}{2}\|w\|^2 + c\sum_{\beta=1}^{m} \varepsilon_\beta$$
$$s.t. \quad y_\beta(w^T x_\beta + g) \geq 1, -\varepsilon_\beta, \quad \beta=1,......,m \quad (1)$$
$$\varepsilon_\beta \geq 0, \quad \beta=1,......,m$$

where $w$ is the normal vector to the hyperplane separating training examples, $|g|/\|w\|$ is the perpendicular distance of the hyperplane from the origin, and $c$ is a penalty parameter. When $c \to \infty$, SVM does not allow any training errors (hard margin classification) and when $0 < c < \infty$, the model allows some training errors, and hence allowing separating nonlinear examples (soft margin). This is a quadratic programming problem which can be solved for the problem's Lagrange duality multiplier $\alpha \in R^3$ as follows:

$$\max_{\alpha} \quad -\frac{1}{2}\sum_{i=1}^{m}\sum_{j=1}^{m}\alpha_i\alpha_j y_i y_j(x_i.x_i) + \sum_{i=1}^{m}\alpha_i \quad (2)$$

$$s.t. \quad 0 \le \alpha \le C, \sum_{i=1}^{m}\alpha_i y_i = 0.$$

In order to solve the duality problem, many analytical approaches are proposed in the literature, depending on the size of dataset and memory limitation considerations. Sequential Minimal Optimization (SMO) [26] is one of the analytic approaches that is used to solve the quadratic programming (QP) problem (2) in many SVM toolboxes such as LIBSVM tool in MATLAB [27]. SMO breaks the QP problem into multiple smaller subproblems, which are then solved analytically. SMO picks two support vectors, finds corresponding Lagrange multipliers and repeats this process until reaching convergence (within a user-defined tolerance) or a maximum number of iterations.

By solving the duality problem (2), the final hyperplane only depends on the support vectors (i.e., sample points that are in the margin) and SVM needs to find only the inner products between the test samples and the support vectors. Fig. 1 shows the support vectors and optimal hyperplane in a separable two-class classification of SVM. In regards to the objective of this paper, Fig. 1 also shows the support vectors and optimal hyperplane to separate outage from operational components based on the associated resiliency index, distance from the center of the hurricane, and the wind speed.

The idea of the maximum-margin hyperplane, which is discussed above, is based on the assumption that training data are linearly separable. To apply SVM to nonlinear data (which often is the case, especially in the case of the hurricane data), kernel methods [25] can be used. The idea of a kernel method (or as sometime called kernel trick) is to map the input space into a linear separable feature space, usually a higher dimension, where the linear classifiers can separate two classes (Fig. 2). Kernel trick simply states that for all $x_1$ and $x_2$ in the input space, a certain function $k(x_1,x_2)$ can be replaced as inner product of $x_1$ and $x_2$ in another space. For example, a Gaussian kernel can be defined as:

$$k(x_i,x_j) = e^{-\frac{1}{2\sigma^2}\|x_i - x_j\|^2} \quad (3)$$

where $\sigma^2$ is the parameter of the kernel defined by the user. In practice, the best kernel is found by experiment while adjusting kernel parameters via a search method to minimize the error on a test set.

B. *Component Features*

A feature, in machine learning, is defined as an individual measurable property of a phenomenon being observed [17]. Selection of discriminating, independent, and informative features plays a critical role in the performance of the classification method. Various features can be defined to determine the state of the components in response to a hurricane strike. In [28], the wind speed and the distance of the each component from the center of the hurricane are proposed as response to a hurricane.

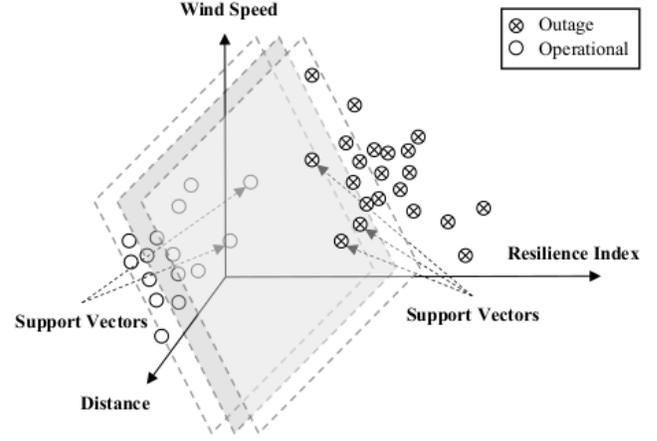

Figure 1. Support vectors and optimal margin in SVM

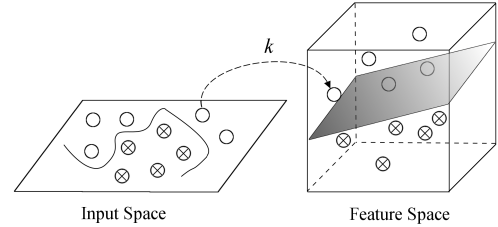

Figure 2. The kernel method in SVM. The linearly inseparable data in a two-dimensional space can be linearly separable in higher dimensions (three dimensions in this figure)

Although these features are obviously adequately informative, they do not provide information about the component itself. Resilience index of components is also an important factor during weather-related events. Similar to [16], we quantify the hurricane resilience of the electric power grid using a probabilistic modeling approach. For the sake of illustration, only the Poisson process model of hurricane occurrence during a given time period along with fragility models are considered in this work. Other factors used in [16] such as DC power flow, power grid restoration and component repair priority are not considered in this index. However, the proposed model is a general framework and can be extended to other resilience indices. Based on this, hurricanes are described by a Poisson process of constant rate $\lambda_h$ such that the time interval between successive hurricane events has an exponential distribution with a probability function of

$$f(t) = \begin{cases} \lambda_h e^{-\lambda_h t} & t \ge 0 \\ 0 & t < 0 \end{cases} \quad (4)$$

Similar to [16] and based on historical data from 1900 to 1999 [29], the annual occurrence rate of hurricanes is considered as $\lambda_h = 1/7$ per year, and the probability of a hurricane belonging to each category is respectively calculated as 0.53, 0.19, 0.15, 0.08, and 0.05. In this paper, we consider resilience index for four components: a) generation units, b) transmission lines, c) distribution lines, and d) substations. For their flexible analytical properties, similar fragility models following a normal distribution, are considered for all four categories with probabilities of low, moderate, severe, and complete. Resilience index is then considered as the average of fragility model and the probability of the hurricane. The category of hurricane, the distance of each component from the center of the hurricane, and the calculated components resilience index are investigated as three main features to predict the state of each component in response to the hurricane.

*C. Evaluation Metrics*

There are various evaluation metrics in the literature to measure the reliability and acceptable performance of a classification method. Accuracy is the most common measure of any classification system which is commonly defined as the number of correct predictions divided by the total number of samples in the test set. Reporting the general accuracy of prediction cannot be sufficient as the number of samples may not balance in the test set. To test the performance of the obtained decision boundary, the $F_1$-Score [30] will be tested on the test historical data defined as:

$$F_1 = \frac{2PR}{P+R} \quad (5)$$

where $P$ is the number of positive predictions divided by the total number of positive class values predicted (i.e., precision), and $R$ is the number of positive predictions divided by the number of positive class values in the test data (i.e., recall). For example, in the case of the outage prediction problem, precision ($P$) is the number of correctly predicted outages divided by the total number of predicted outages, and recall ($R$) is the number of correctly predicted outages divided by the total number of actual outages. The $F_1$-Score will be a value ranging from 0 to 1, where higher values represent a higher predictive power as a measure of acceptable performance of the obtained decision boundary.

To evaluate the performance of the classifier, usually a subset of the historical dataset is reserved as holdout sample for model validation. $k$-fold cross-validation is a common validation technique for assessing the results of a classification system and evaluating how well it can generalize on a dataset [31]. In $k$-fold cross-validation, the dataset is randomly partitioned into $k$ equal sized subsamples. A single subsample is reserved as the validation/test set, and the other $k−1$ subsamples are used as training data for the model. This process is iterated for $k$ times (the number folds), where each of the $k$ subsamples is used only once for the validation. The $k$ results from the folds are accordingly averaged to obtain a single estimation.

## III. NUMERICAL SIMULATION

Scarcity of readily available datasets still remains a challenge for research community and industry practitioners. However, the limited historical data on past extreme hurricanes at the component granularity level shall not preclude methodological developments in critical areas including in machine learning systems. Therefore, in this paper, a synthetic set of 1000 sample data is generated to train the SVM model, considering half of the samples in outage state and the other half in the operational state. The generated samples follow a normal distribution function of one-minute sustained wind speed of different Saffir-Simpson Hurricane Scale categories with a small Gaussian noise. The features are normalized to [0,1] based on the maximum considered values of wind speed and distance. Fig. 3 shows the generated synthetic data.

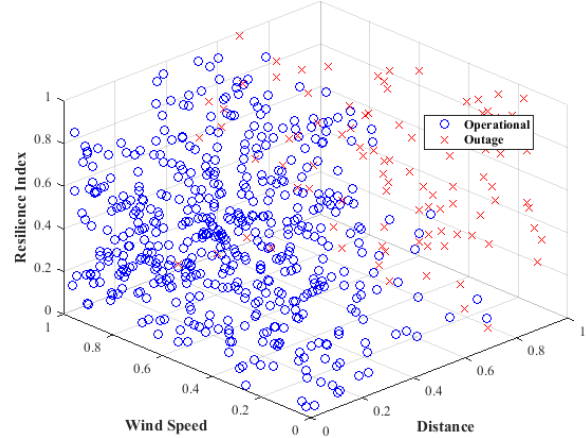

Figure 3. Generated synthetic data for SVM training and validaton

A $k$-fold cross validation ($k$=5) is performed to measure the performance of the proposed model. Different kernels (linear, polynomial Quadratic, Cubic, and Gaussian) with various penalty parameters (c=0.01, 0.1, 1, 10, 100) are examined. Since the considered dataset is relatively small, an off-the-shelf SVM model implemented in LibSVM [27] is used in this paper. In the proposed work, the SMO tolerance for convergence is set to 1e-3 and the maximum number of iterations is set to a large value (15000 iterations). In practice, since the considered dataset is relatively small, it converges in about 350 iterations for different folds. Table I shows the average $F_1$-Score for various penalty parameters and kernel shapes. As it is shown, SVM with Gaussian kernel and $c$=1 offers the best performance among other settings.

A third order polynomial logistic regression model is also trained and examined in the same fashion (i.e., $k$-fold cross-validation with $k$=5) to predict the component outages. Table II compares evaluation metrics of SVM with different kernels (using penalty parameter $c$=1) and a third order polynomial logistic regression model. As shown, among the trained models, Gaussian kernel SVM had the best overall classification accuracy with a precision of 0.893, a recall of 0.826, and overall $F_1$-Score of 0.858. Comparing the result of logistic regression with the proposed SVM indicates that the proposed SVM approach has a better performance in both accuracy and $F_1$-Score.

Table III shows confusion matrix of predicting components as operational and outage using Gaussian kernel SVM. The proposed model can predict outage and operational states with the accuracy of 90.2% and 82.6%, respectively.

TABLE I. AVERAGE F1-SCORE OF SVM WITH VARIOUS PANEALTY PARAMETERS "C" AND KERNELS USING 5-FOLD CROSS-VALIDATION

| Kernel | c=0.1 | c=1 | c=10 | c=100 |
|---|---|---|---|---|
| Linear | 0.845 | 0.845 | 0.846 | 0.846 |
| Quadratic | 0.858 | 0.856 | 0.855 | 0.857 |
| Cubic | 0.855 | 0.854 | 0.840 | 0.754 |
| Gaussian | 0.857 | **0.858** | 0.850 | 0.847 |

TABLE II. COMPARISON OF THE PERFORMANCE OF SVM WITH VARIOUS KERNELS AND THE LOGISTIC REGRESSION METHOD.

|  | Accuracy | Precision | Recall | F1-Score |
|---|---|---|---|---|
| Linear SVM | 0.847 | 0.853 | 0.838 | 0.845 |
| Quadratic SVM | 0.863 | 0.898 | 0.818 | 0.856 |
| Cubic SVM | 0.861 | 0.896 | 0.816 | 0.854 |
| Gaussian SVM | **0.864** | 0.893 | 0.826 | 0.858 |
| Logistic Reg. | 0.809 | 0.815 | 0.798 | 0.806 |

TABLE III. CONFUSION MATRIX OF CLASSIFYING SYSTEM COMPONENTS USING GAUSSIAN KERNEL SVM (NUMBER OF SAMPLES AND PERCENTAGE)

|  |  | Predicted | |
|---|---|---|---|
|  |  | Normal | Outage |
| Actual | Normal | 451 (90.2%) | 49 (9.8%) |
| | Outage | 87 (17.4%) | 413 (82.6%) |

## IV. CONCLUSION

Prediction of a component state in response to an extreme event is a challenging task in practice. In this paper, a three dimensional SVM was proposed to categorize system components into two classes of damaged and operational in response to an upcoming hurricane. The proposed SVM was trained on historical data with three features related to each grid component—i.e., the resilience index, the distance of the component from the center of the hurricane, and the category of the hurricane (the wind speed). A synthetic set of data was generated to train the SVM, as the publicly available data on the impact of hurricanes on power grid components is limited. Simulation results showed the effectiveness of the proposed model compared to the results obtained from Logistic Regression, as a popular benchmark for two-class classification problem, and further demonstrated its acceptable performance in reaching high accuracy estimations.